# Genuine onion: Simple, Fast, Flexible, and Cheap Website Authentication

Paul Syverson
U.S. Naval Research Laboratory
paul.syverson@nrl.navy.mil

Griffin Boyce
Open Internet Tools Project
griffin@cryptolab.net

*Abstract*—Tor is a communications infrastructure widely used for unfettered and anonymous access to Internet websites. Tor is also used to access sites on the .onion virtual domain. The focus of .onion use and discussion has traditionally been on the offering of hidden services, services that separate their reachability from the identification of their IP addresses. We argue that Tor's .onion system can be used to provide an entirely separate benefit: basic website authentication. We also argue that not only can onionsites provide website authentication, but doing so is easy, fast, cheap, flexible and secure when compared to alternatives such as the standard use of TLS with certificates.

## I. INTRODUCTION

Tor is a widely popular communications infrastructure for anonymous communication. Millions use its thousands of relays for unfettered traffic-secure access to the Internet. The vast majority of Tor traffic by bandwidth (over 99% at our last check) is on circuits connecting Tor clients to servers that are otherwise accessible. Tor also provides protocols for connecting to services on the pseudo-top-level domain .onion, which are only accessible via Tor. In this paper we explore using Tor's .onion infrastructure so that individuals operating a website can create authentication, integrity and other guarantees more simply, easily, fully, cheaply, and flexibly than by relying only on standard protocols and authorities.

Tor's onionsites have been advocated since their introduction as a way to protect network location information for servers not just clients [1]. [1] Discussion in the popular press, as well as research to date, has focused almost exclusively on location hiding and associated properties provided by onionsites and the protocols to interact with them. Indeed, these are generally referred to collectively as *Tor Hidden Services* in the research literature and as the *Dark Web* in the popular press. (Although so many importantly distinct things are often subsumed and run together under 'Dark Web' as to rob the term of clear significance, other than as a caution flag for the hidden incoherence that surrounds most occasions of its use.)

Our intent is to challenge the narrowness of this view of onionsites. In particular we will discuss security protections they readily facilitate that are largely orthogonal to hiding server location. We hope by the end of this paper the reader will agree that they should be called *onion services* or in any case something that is more properly inclusive of the variety of security properties they offer.

---

[1] Such advocacy actually predates their introduction inasmuch as the same was said for web servers contacted by reply onions [2], See also [3] for description of an implemented predecessor to Tor's hidden services.

## II. BRIEF BACKGROUND ON TOR AND ONION SERVICES

We sketch out minimal basics of Tor onion services. For more detailed descriptions see the Tor design paper or other documentation at the Tor website [4]. For a high-level graphical description of onion services see [5]. For a more up to date, and much more technical, description of onion services protocols see the Tor Rendezvous Specification [6].

Tor clients randomly select three of the roughly 6000 relays comprising the current Tor network, and create a cryptographic circuit through these to connect to Internet services. Since only the first relay in the circuit sees the IP address of the client and only the last (exit) relay sees the IP address of the destination, this technique separates identification from routing. To offer an onion service, a web (or other) server creates Tor circuits to multiple *Introduction Points* that await connection attempts from clients. A user wishing to connect to a particular onion service uses the onion address to look up these Introduction Points in a directory system. In a successful interaction, the client and onionsite then both create Tor circuits to a client-selected *Rendezvous Point*. The Rendezvous Point mates their circuits together, and they can then interact as ordinary client and server of a web connection over this rendezvous circuit.

Since the onionsite only communicates over Tor circuits it creates, this protocol hides its network location, the feature that gives it the name 'hidden service'. But, there are other important features to the .onion system, notably self-authentication. The onion address is actually the hash of the public key of the onionsite. For example, if one wishes to connect to the DuckDuckGo search engine's onion service, the address is 3g2upl4pq6kufc4m.onion. The Tor client recognizes this as an onion address and thus knows to use the above protocol rather than attempting to pass the address through a Tor circuit for DNS resolution at the exit. The public key corresponds to the key that signs the list of Introduction Points and other service descriptor information provided by the directory system. In this way, onion addresses are self-authenticating.

## III. KNOWING TO WHICH SELF TO BE TRUE

Of course this authentication only binds the service descriptor information to the 3g2upl4pq6kufc4m.onion address. What a user would like to be assured of is that s/he is reaching DuckDuckGo. Presumably the user wants the search results DuckDucGo offers and not what might be returned by some other, possibly malicious, server. In addition to the integrity guarantee, the user relies on authentication so that queries are revealed only to DuckDuckGo and not to others.

The onion address by itself does not offer this. Making use of the traditional web trust infrastructure, DuckDuckGo and Facebook offer certificates for their onion addresses issued by DigiCert. This helps ensure that users are not misled by onionsites purporting to be official.

Though cryptographic binding is essential to the technical mechanisms of trust, users also rely on human-readable familiarity, for example, that the browser graphically indicates s/he has made a certified encrypted connection as a result of typing facebook.com into the browser. To some extent, it is at least possible to make use of this in .onion space. By generating many keys whose hash had 'facebook' as initial string and then looking among the full hashes for an adequately felicitous result, Facebook was able to obtain facebookcorewwwi.onion for its address. Whatever its value for Facebook, this is clearly not something that will work widely, as it is difficult to generate custom addresses in this way.

Why not just obtain certificates from traditional authorities as DuckDuckGo and Facebook have done? For many server operators, getting even a basic server certificate is just too much of a hassle. The application process can be confusing. It usually costs money. It's tricky to install correctly. It's a pain to update. These are not original observations. Indeed that description is actually a quote from the blog of Let's Encrypt, a new certificate authority dedicated, among other things, to making TLS certification free and automatic [7].

Using the existing X.509 system, setting up a certificate can take hours or even days. In cases where the website is operated by a collective or organization, SSL/TLS certificates have been known to take months, due to questions around ownership and authorization. This time cost is in addition to the monetary cost of the certificate, if any. In contrast, setting up an onionsite takes a few minutes and costs nothing. Once Tor is installed, you simply add two lines to your torrc file and start Tor. The Tor Project also provides a brief page with additional tips and advanced options [8]. Even when the process of learning how to create a PGP key and signature is taken into account, the time investment is dramatically less than with the current X.509 public-key infrastructure.

As of this writing, Let's Encrypt services are still a few months away. Should they be willing to offer certificates for onion domains, using Let's Encrypt could be an easy way for onionsite operators to take advantage of the traditional certification infrastructure. Traditional certificates are not without problems, however. The nature of the trust hierarchy is opaque to direct usage, and the sheer number of trusted authorities is large enough to be of concern. In particular, there have been numerous cases of man-in-the-middle (MitM) attacks through certificate manipulation, as well as hacking of certificate authorities or certificate validation software leading to use of fraudulent certificates for some of the most popular websites [9].

EFF's SSL Observatory [10] monitors for such problems and documents their occurrence. Google's Certificate Transparency Effort [11] is similar but broader, adding (amongst other things) append-only signed public logs that make certificate shenanigans all the harder to bring off undetectably.

Rather than simply monitor and flag certificate authority problems, the Perspectives Project [12] strives to provide end users with control over the trust they place in website certificates. Instead of trusting anointed CAs, semi-trusted network notaries probe network services and build a record of public keys those services have used over time, somewhat similar to the approach of certificate transparency. Users can choose which notaries they wish to trust, and clients encountering unfamiliar public keys will query notaries for a history of keys used by a service [13]. This is especially intended to enhance trust on first use (tofu) authentication, although it can also supplement traditional CA based PKI security.

## IV. OUR ONIONS OURSELVES

As noted, onionsites already provide a self-authenticated binding of public key to onion address but do not bind that public key to something recognizably associated with that site. Even for hidden service applications, it might still be desired to connect the onionsite to some pseudonymous reputation. We will view location hiding as largely orthogonal. We seek a solution that will work for all kinds of sites, but we are especially interested in providing authentication for small and/or short-lived websites, e.g., personal web pages, hometown sports teams, sites for local one-time events, small businesses, municipal election campaigns, etc. Though not such large targets as more popular long-lived sites, they are still subject to controversy and have been subject to many of the same sorts of attacks as more well-known sites. They might also not be the target of attacks but simply collateral victims.

Some users of this kind may not even have Internet accounts that allow them to set up servers. Onionsites are compliant with such a limitation since they actually only make outbound client connections. As a related example of an existing usage, many people administering systems behind restrictive firewalls that only permit outbound connections currently use onion services to administer their systems. Even if the user has an Internet account that permits setting up a web page, HTTPS may not be available from that provider or only available for an additional fee.

We are primarily focused on improvements to authentication using onionsites and thus mostly leave properties of network location hiding aside as orthogonal to our goals. They can be complementary, however. Authenticated hidden services are an appealing option for those who'd like to secure their onionsites for personal use. Unlike traditional websites that appear online prior to authentication, users lacking authentication information for such a site will not be easily able to determine that it even exists, nor will they be able to probe it for vulnerabilities. These qualities make an ideal environment for operating a personal cloud service. With privacy and cost in mind, many people are operating their own cloud infrastructure to store files and calendar entries using open-source systems such as Cozy and OwnCloud [14]. Another common use of authenticated hidden services is as a personal RSS reader, as onionsites ensure some level of feed integrity (particularly important when fetching news feeds that do not utilize TLS).

Of course one can always create a Facebook page or something similar that is protected by HTTPS and TLS certificates, and this is often done. But this makes the service dependent on the reputation, trust, policies, and protections of such a host, not to mention the dynamics thereof, rather than allowing the user to readily understand and control these aspects of his own service. Also, Facebook policy requires identification of the person providing the site, while we would prefer to leave this as simply a separate issue.

A very simple way to add binding of the onionsite public key to a known entity using widely available mechanisms is to provide a signature on the onion address. We envision a PGP/GPG signature, but it could be an X.509 signature (or other as we discuss below). The signed text can simply be included on the onionsite, making it self-authenticating in this sense as well. The trust in the authentication will then be whatever trust is associated with the public key that does the signing. Such techniques are already used for signing code. For example, the Tor Project offers signatures on all source and binaries it makes available for download.

If the signer wishes to post the signed onion address to a public site such as her Facebook page, she can do this also. (An advantage of doing so will be discussed below.) Indeed, a useful public site for doing this would be an unauthenticated version of the same exact service as the one being offered at the onionsite. The unauthenticated version and the onionsite version should both contain a signed pointer to both versions. It is then easy for anyone who desires to check their association. For example, we have made an authenticated version of http://cupcakebridge.com available at http://eynfqhbaq5yecx6s.onion.

Why even bother with the non-onionsite version? There are several reasons. First, this allows for a binding of the public domain name to the onionsite. As mentioned, onion addresses are inherently not humanly meaningful, which can lead to confusion among end-users. To get the entirety of a specific domain name of choice is also technologically infeasible, as onion addresses are randomly-generated alphanumeric strings of 16 digits. The signed-onion technique allows someone to choose and retain a desired domain name for the site, while still being able to offer an authenticated and integrity protected version easily. This also illustrates one of the benefits of using GPG or similar signatures. If the authentication simply showed that the same party that provided the not-secured site provided the onionsite, an attacker could set up an altered version, employ usual techniques to hijack the not-secure site, and offer a self-authenticated onionsite that matched the hijacked site. To do this undetectably against the GPG-signed onionsite would require subversion of the trust in the GPG key. A concern with using GPG signatures is that users may not be familiar with them, have appropriate trust in the key, or bother to check the signature and the trust in the key. We will address these below.

In addition, many intended users of the site may not have Tor installed. Though installation is a simple point-and-click download, many may be disinclined against even this small effort. The onionsite would still be available via Tor2web, a website that proxies connections from non-Tor clients to onionsites [15]. To connect to an onionsite, one enters a URL such as the following for reaching DuckDuckGo's onionsite via Tor2web: https://3g2upl4pq6kufc4m.tor2web.org/. The Tor2web site explicitly states that "Tor2web only protects publishers, not readers." This is because the client connects to Tor2web over a direct TLS connection rather than via Tor, as would be the case of someone connecting to 3g2upl4pq6kufc4m.onion via the Tor Browser. For our purposes, authentication of the onionsite in this case is limited to the trust in authentication of this TLS connection (and trust in Tor2web itself) regardless of the trust in the GPG signature. Thus, no significant usability limitation arises from using other browsers, but security is significantly downgraded for the reasons already mentioned and by various MitM possibilities raised below.

Finally, traditional search and indexing engines such as Google do not generally reflect links to onionsites. The search engine, ahmia.fi [16], is limited to onionsites and known primarily to those familiar with them. Ahmia creator Juha Nurmi has, however, agreed to incorporate linking of onion and clearnet addresses into Ahmia, together with the GPG signatures binding that linking. He has also suggested to us that Ahmia could automatically test the signatures and check the clearnet and onion sites. Thus, a user who trusts Ahmia (and her connection to Ahmia) on this can verify that a pair of websites is operated by the same party, even if personally inexpert with manual PGP verification. Crawling and indexing of onionsites is also in its infancy and can thus not be expected to be as appropriately representative as the much more mature indexing of the surface web by Google and similar sites.

V. USABILITY, CONVENIENCE, AND SECURITY

As most onionsite visitors use the Tor Browser, deployment and debugging of hidden services can be faster than their clearnet counterparts because there is only one browser to test, with only minor variation across users. Website operators can assume that users do not have AdBlock or other browser extensions that may impact how content is displayed. However, plugins that may mitigate Tor Browser's privacy protections, such as Java and Flash, have been disabled by default. Many privacy-conscious users do enable the NoScript extension to block javascript as well. Despite this, rich content such as video, audio, and interactive storytelling are still available for designers willing to use HTML5 and CSS3.

What we have described so far implies a relatively manual authentication of PGP/GPG signatures. It would be natural and straightforward to create a plugin that verifies the signature and provides different indications to the user depending on the trust in it. There are already related tools, e.g., Monkeysphere, a Firefox plugin that uses the PGP trust infrastructure for validation only when the browser does not default accept the TLS certificate validation [17]. A simpler plugin could also just check the planned Ahmia validation mentioned above.

Our approach naturally complements Perspectives and similar endeavors. Perspectives offers an improvement against certificate based MitM attacks. But if a site is newly available, the onionsite can still be trusted to be bound to the owner of the PGP signature. A new self-signed certificate, on the other hand, will have little or no Perspectives history, and users are reduced to a tofu decision. Also, Perspectives notaries largely function as detectors of certificate misbehavior over time. This is useful but cannot detect static misuse of certificates. For example, consider a typo-squatting site that uses a self-signed certificate to pass through connections to the site on which it is squatting, but does not misbehave or alter its key. Perspectives will not reflect anything wrong with such a site, whereas our approach will presumably not give the site a high degree of trust unless the squatter has that trust exogenously.

Our approach also can be used (at least in manual form) right now by website operators. It would benefit from usability developments and simplification, and it can complement other approaches. It does not, however, rely fundamentally on the deployment and continued commitment to new infrastructure that is specific to it. It can instead rely on whatever authentication infrastructure might be popular and likely to be maintained for independent reasons, rather than needing to grow and maintain interest in its approach.

Convergence takes a similar view and *is* automated and deployed (but currently is designed for existing TLS signatures). Extending Perspectives, Convergence notaries can use additional strategies beyond network perspective to create trust [18]. For example, in addition to perspective notaries, one can choose to place some trust in tradional CAs, or DNSSEC authorities, or anyone you like. Also, instead of trusting an authority or class of authorities essentially forever as is the case for existing approaches, Convergence more straightforwardly allows one to add or remove notaries. Convergence is available in Firefox plugins. It would be interesting to investigate a Convergence-like addition to the Tor Browser that works with onion addresses and PGP/GPG signatures.

In the PGP web of trust, signature authority is built up in a decentralized manner from direct personal connections and introductions. This more naturally fits with many of the kinds of websites that we have suggested could most benefit from our approach, for which local or personal trust relationships are important [19]. The X.509 trust model currently in general use to support TLS certificates is by contrast primarily a hierrarchical centralized chain of trust delegated down from some ultimate national or global corporate trust anchor.

PGP and its successors also remain less familiar than TLS. Although popular familiarity is not so much with TLS as with interfaces that tell the user little more than whether or not TLS is in operation at all, which is essentially as it should be for usable security. As noted, similar interfaces for PGP have been designed but have not yet received the extensive development of TLS interfaces, unsurprising given the fundamental role of TLS in global ecommerce. For those who do not rely on the social or local protections of PGP's web of trust, TLS certificates are likely to remain the primary ground of linking public, human-readable domain names to the signatures authenticating websites. Assuming Let's Encrypt is successful, and no-additional-cost HTTPS is increasingly available from ISPs, we can also envisage incorporation of TLS with onionsites for even the "everyman" users described above. To the extent that this incorporation follows what Facebook and DuckDuckGo have done, the strength of trust in the authentication is limited by the trust in the TLS certificate. Certificate transparency and the like will help here, but the self-authentication of onion addresses can also add to this trust, and again in a way more directly under website owner control.

Alternatively something like Convergence could integrate the advantages from onionsite self-authentication that our approach provides with the TLS signatures familiar to existing web browsers. But it would not need to rely on existing CAs, which is indeed a primary goal of Convergence. This seems a promising direction, although not as immediately usable as the manual check we describe. It requires both a web of trust for TLS signature keys and integration with onionsites to adequately develop.

Also, unlike conventional web URLs, onion addresses are inextricably connected to the site authentication key. This means that if one has publicized the onion address, e.g., through blogs, twitter, or Facebook, people following those address links will not be vulnerable to hijack or MitM by the subverted CA the way they would be by a link to a regular URL. This significantly raises the bar on the hijacker fairly automatically and easily. Further, non-CA-based MitM techniques such as forcing the site to fall back to a non-SSL version (e.g., SSLStrip) or to use a weaker cipher to communicate (e.g, BEAST and FREAK) are also not possible as the address and key are inextricably linked and generated using a strong cipher. And, if onionsite private keys were to sign not just the Introduction Points and other elements currently stored in the .onion directory system, but also the TLS certificate, onion keys would bind the TLS certificate to the site as well as vice versa.

VI. CONCLUSION

In this paper we have described how Tor's onion services can be used not for the usual stated purpose of hiding server network location, but for website authentication. We have also argued that onion services offer users, a simple, effective, cheap and flexible means of authentication with security advantages not provided by existing approaches. We hope people will find this useful and begin employing it. We also hope our expanded view of the possibilities created by Tor's onion services will encourage others to explore this fascinating system for other interesting properties and applications.